\documentstyle[amssymb,eqsecnum,aps]{revtex}

\begin{document}
\draft
\title{Quantum Theory of Generation of Coherent X-Ray in a Wiggler }
\author{H.K. Avetissian, T. R. Hovhannisyan, G.F. Mkrtchian}
\address{Department of Quantum Electronics, Plasma Physics Laboratory, Yerevan State 
University \\
1 A. Manukian, Yerevan 375025, Armenia\\
E-mail: avetissian@ysu.am }
\maketitle

\begin{abstract}
In this paper we present the nonlinear quantum theory of X-Ray FEL in a
wiggler. We present the solution of the Dirac equation in a space periodic
strong magnetic field, which describes the quantum dynamics of a single
electron in a wiggler. With the help of obtained wave function of an
electron the self-consistent set of the Maxwell and relativistic quantum
kinetic equations is obtained. Then, the process of coherent X-ray radiation
generation in nonlinear quantum regime, when the photon energy is larger
than resonance widths due to energetic/angular spreads of an electron beam
and resonance width caused by finite interaction length, is investigated.
\end{abstract}

\pacs{PACS number(s):  41.60.Cr, 42.55.Vc, 42.50.Fx}

\section{Introduction}

Many types of stimulated radiation sources have been suggested over the
years to achieve lasing in shortwave domain. A promising way for realization
of X-ray laser is connected with the version of free electron laser (FEL) 
\cite{FEL}. The main advantage of FEL lies in the fact that the emission
frequency is continuously Doppler upshifted by several orders ($\sim \gamma
_L^2,$ $\gamma _L$ being Lorenz factor) with respect to the frequency of the
pump field.{\it \ }In particular, the X-ray FEL schemes have been proposed
based on the coherent accumulation of ultrarelativistic electron beam
radiation in undulator and at Compton-backscattering, channeling, transition
and diffraction radiations \cite{List}. At the present, however, among these
versions the undulator scheme is actually developed and the first lasing has
been carried out in this system \cite{Smith}. Although the amplifying
frequencies are still far from X-ray, the main hopes for an efficient X-ray
FEL remain connected with the undulator scheme. The absence of the
normal-incidence mirrors of high reflectivity at the short wavelengths in
order of X-ray practically excepts a resonator scheme of radiation
generation. In this case it is necessary to implement a single pass high
gain FEL. The most attractive scheme, which is presently considered, is the
Self Amplified Spontaneous Emission \cite{SASE}, where the spontaneous
undulator radiation from the first part of an undulator is used as an input
signal to the downstream part. For this purpose two international projects 
\cite{Int} TESLA and LCLS being currently implemented.

In the more conventional undulator devices to achieve X-ray frequencies
domain one should increase the electrons energies up to several $GeV$, which
in turn significantly reduces the small-signal gain ($\sim \gamma _L^{-3}$ ).

In contrast with conventional laser devices on atomic systems FEL is usually
reckoned as a classical device. But this is not universal property of FEL as
in some cases the quantum effects may play significant role, especially for
X-ray frequencies. In the quantum description \cite{qfel}, the resonant
momenta of an electron for the emission $p_e$ and absorption\ $p_a$ are
different due to a quantum recoil. The probability distributions of emission
and absorption are centered at $p_e$\ and $p_a$\ respectively\ and when
these distributions much narrower than the spread of electron beam
distributions $f(p)$, the small-signal gain is proportional to the so called
''population inversion'' $f(p_e)-f(p_a)$. In the quasiclassical limit when
amplifying photon energy $\hbar \omega $ fulfils condition {\em \ } 
\begin{equation}
\hbar \omega <<\max \left\{ \Delta \varepsilon _\gamma ,\Delta \varepsilon
_\vartheta ,\Delta \varepsilon _L\right\}  \label{i1}
\end{equation}
where $\Delta \varepsilon _\gamma $, $\Delta \varepsilon _\vartheta $, are
the resonance widths due to energetic and angular spreads, and $\Delta
\varepsilon _L$ resonance width caused by finite interaction length the
quantum expression for the gain coincides with its classical counterpart
being antisymmetric about the classical resonant momentum $p_{c=}(p_e+p_a)/2$
and proportional to the derivative of the momentum distribution function $%
df(p)/dp$ at $p=$ $p_c$. Resulting gain takes place only if the initial
momentum distribution is centered above $p_c$\ as the electrons whose
momenta are above $p_c$\ contribute on average to the small-signal gain, and
those whose momenta are below $p_c$\ contribute on average to the
corresponding loss. This severely limits the FEL gain performance at short
wavelengths.

The efficiency of FEL at short wavelengths can be significantly increased in
the quantum regime of generation: 
\begin{equation}
\hbar \omega \gtrsim \max \left\{ \Delta \varepsilon _\gamma ,\Delta
\varepsilon _\vartheta ,\Delta \varepsilon _L\right\} ,  \label{i2}
\end{equation}
In this case the absorption and emission lineshapes are separated and the
simultaneous absorption of probe wave is excluded. To achieve the condition (%
\ref{i2}) for FEL operation is problematic as it presumes severe
restrictions on the beam spreads. In any case, it may be satisfied at the
emission of hard X-ray quanta.

In this work the scheme of X-ray coherent radiation generation in nonlinear
quantum regime by means of relativistic high density electron beam in
wiggler is investigated. The possibility to achieve quantum regime of FEL at
high harmonics of Doppler-shifted ''wiggler frequencies'' is treated. The
consideration is based on the self-consistent set of the Maxwell and
relativistic quantum kinetic equations. Because of nonequdistantness of the
energy-momentum levels the probe wave resonantly couples only two electrons
states in wiggler and the coupled equations are solved in the slow varying
envelope approximation.

This paper is organized as follows. In Section II we obtain the wave
function of an electron in a wiggler. Section III describes our model with
the self-consistent set of equations. The particular solutions of
self-consistent set of equations for the X-ray generation are discussed in
Sec.\ IV. Finally, conclusions are given in Sec.\ V.

\section{Wave Function of an Electron in a Wiggler}

For the quantum description of FEL dynamics we will need the wave function
of an electron in a wiggler. We will consider as linear (LW) as well as
helical Wigglers (HW). Here and in what follows for the four-component
vectors we have chosen the following metrics $a\equiv a^\mu =(a_0,{\bf a})$
and $ax$ is the relativistic scalar product: 
\[
ax\equiv a^\mu x_\mu =a_0x_0-{\bf ax.} 
\]
To describe the magnetic field of a Wiggler we will choose the following
four-vector potential

\begin{equation}
A_H^\mu =(0,{\bf A}_H),  \label{W1}
\end{equation}
where 
\begin{equation}
{\bf A}_H=(A_0\cos (-k_0x),gA_0\sin (-k_0x),0),  \label{W2}
\end{equation}
$x=(ct,{\bf r})$ is the four-component radius vector and

\begin{equation}
k_0\equiv (0,{\bf k}_0)=(0,0,0,\frac{2\pi }\ell ),  \label{k}
\end{equation}
with the Wiggler step-$\ell $. In (\ref{W2}) $g=\pm 1$ correspond to HW,
while $g=0$ corresponds to LW.

The dynamics of an electron in Wiggler can be described by the Dirac
equation, which in the quadratic form \cite{Lan}, taking into account (\ref
{W2}), is the following 
\begin{equation}
\left\{ \left( i\hbar \partial _\mu +\frac ecA_{H\mu }\right) ^2-m^2c^2+%
\frac{e\hbar }c\widehat{{\bf \Sigma }}{\bf H}\right\} \psi =0,  \label{4}
\end{equation}
where $\hbar $ is the Plank constant $m$ and $e$ are the particle mass and
charge respectively, $c$ is the light speed in vacuum and $\partial _\mu
\equiv \partial /\partial x^\mu $ denotes the first derivative of a function
with respect to four-component radius vector $x$. Here 
\begin{equation}
\widehat{{\bf \Sigma }}=\left( 
\begin{array}{cc}
\widehat{{\bf \sigma }} & 0 \\ 
0 & \widehat{{\bf \sigma }}
\end{array}
\right)  \label{5}
\end{equation}
is the spin operator with the $\widehat{{\bf \sigma }}$ Pauli matrices and 
\begin{equation}
{\bf H}=rot{\bf A}_H  \label{6}
\end{equation}
is the magnetic field of a Wiggler.

As the magnetic field depends only on the $\tau =-k_0x={\bf k}_0{\bf r}$
then raising from the symmetry, we seek a solution of Eq. (\ref{4}) in the
form 
\begin{equation}
\psi (x)=f(\tau )\exp \left[ -\frac i\hbar px\right] ,  \label{7}
\end{equation}
where $p=(\varepsilon /c,{\bf p})$ is the four-momentum of a free Dirac
particle.

To solve Eq. (\ref{4}) we will consider $f(\tau )$ as a slowly varying
bispinor function of $\tau $ (in the scale of $pk_0/(\hbar k_0^2$) and
neglect the second derivative compared with the first order (see condition (%
\ref{con})). So from (\ref{4}) and (\ref{7}) for $f(\tau )$ we will have the
following equation:

\begin{equation}
-2i\hbar (pk_0)\frac{df(\tau )}{d\tau }+\left\{ -\frac{2e(pA_H)}c+\frac{%
e^2A_H^2}{c^2}+\frac{e\hbar }c\widehat{{\bf \Sigma }}{\bf H}\right\} f(\tau
)=0.  \label{8}
\end{equation}
The solution of Eq. (\ref{8}) we can write in the operator form

\begin{equation}
f(\tau )=\exp \left\{ \frac i{\hbar c(pk_0)}\int_0^\tau \left( e(pA_H)-\frac %
e{2c}A_H^2\right) d\tau ^{\prime }-\frac{ie}{2c(pk_0)}\widehat{{\bf \Sigma }}%
\left[ {\bf k}_0{\bf A}_H\right] ,\right\} w(p)  \label{9}
\end{equation}
where $w(p)$ is the arbitrary bispinor amplitude. Then taking into account
the property of spin operator 
\[
\exp \left[ \widehat{{\bf \Sigma }}{\bf a}\right] =\frac 12(\exp (a)+\exp
(-a))+\widehat{{\bf \Sigma }}{\bf a}\frac 1{2a}(\exp (a)-\exp (-a)) 
\]
and putting condition $a<<1,$ which in our case restricts the magnetic field
strength by the condition

\begin{equation}
K=\frac{eA_0}{mc^2}=\frac{eH_0\ell }{2\pi mc^2}<<\gamma _L.  \label{con}
\end{equation}
Here $K$ is the so called Wiggler parameter.

Hence, for the wave function we will have the following expression 
\begin{equation}
\psi (x)=\left[ 1+\frac{e\widehat{k}_0\widehat{A}_H}{2c(k_0p)}\right]
w(p)\exp \left[ -\frac i\hbar \left\{ px-\frac i{\hbar c(pk_0)}\int_0^\tau
\left( e(pA_H)-\frac e{2c}A_H^2\right) d\tau ^{\prime }\right\} \right] .
\label{10}
\end{equation}
Here we have introduced the following notation $\widehat{a}=a^\mu \gamma
_\mu $, where $\gamma ^\mu =(\gamma _0,{\bf \gamma })$ are Dirac matrices.
Note that (\ref{con}) is also the condition of slowly varying function $%
f(\tau )$ over $\tau $. The wave function (\ref{8}) is an analogy of Volkov
wave function \cite{Lan}. The main difference in this case is that $%
k_0^2\neq 0$ but taking into account (\ref{con}) we can neglect the terms
which come from $k_0^2\neq 0$ (this will be more evident in the
Waizs\"{a}cker{\em -}Williams\ approach, when in the frame connected with
electron the wiggler field is well enough described by a plane wave one).

Making integration in the (\ref{10}), taking into account (\ref{W2}), for
the wave function we will have 
\[
\left| {\bf q},\sigma \right\rangle =\left[ 1+\frac{e\widehat{k}_0\widehat{A}%
_H}{2c(k_0p)}\right] \frac{u_\sigma (p)}{\sqrt{2q_0}} 
\]
\begin{equation}
\times \exp \left[ -\frac i\hbar \left\{ qx-\frac{eA_0}{c(pk_0)}(p_x\sin
(-k_0x)-gp_y\cos (-k_0x))-\frac{e^2A_0^2}{8c^2(pk_0)}(1-g^2)\sin
(-2k_0x)\right\} \right]  \label{11}
\end{equation}
where by further analogy with Volkov state we have introduced the
quasimomentum

\begin{equation}
q=p+k_0\frac{m^2c^2}{4k_0\cdot p}(1+g^2)K^2;  \label{q}
\end{equation}
and for arbitrary bispinor we have put 
\[
w(p)=\frac{u_\sigma (p)}{\sqrt{2\varepsilon _0}}, 
\]
where $u_\sigma (p)$ is the bispinor amplitude of a free Dirac particle with
polarization $\sigma _{.}$ It is assumed that 
\[
\overline{u}u=2mc^2\text{,} 
\]
where $\overline{u}=u^{\dagger }\gamma _0$; $u^{\dagger }$ denotes the
transposition and complex conjugation of $u$ (in what follows we will put
the volume of the periodicity $V=1$ ).

So the state of the particle in Wiggler (\ref{11}) is characterized by the
quasimomentum and polarization $\sigma $. The wave function (\ref{11}) is
normalized by the condition

\[
\left\langle {\bf q}^{\prime },\sigma ^{\prime }\mid \mid {\bf q},\sigma
\right\rangle \text{ }=\delta _{{\bf q},{\bf q^{\prime }}}\delta _{\sigma
,\sigma ^{\prime }}\text{,} 
\]
where $\delta _{\mu \mu ^{^{\prime }}}$ is the Kronecker symbol.

\section{Self-consistent set of the Maxwell and Relativistic quantum kinetic
equations.}

In this section we will consider the quantum kinetic equation for a spinor
particles interacting with the classical probe electromagnetic (EM) wave in
a Wiggler.

We assume the probe EM wave to be linearly polarized with the carrier
frequency $\omega $ and four-vector potential 
\begin{equation}
A_w=e_1\left\{ A_e(t,{\bf r})e^{ikx}+k.c\right\} ,  \label{12}
\end{equation}
where $A_e(t,{\bf r})$ is a slow varying envelope, $k=(\frac \omega c,{\bf k}%
)$ is the four-wave vector and $e_1$ is the unit polarization four vector $%
e_1k=0$.

Rising from the second quantization formalism, the Hamiltonian is

\begin{equation}
\widehat{H}=\int \widehat{\Psi }^{+}\widehat{H}_0\widehat{\Psi }d{\bf r+}%
\widehat{H}_{int}  \label{h1}
\end{equation}
where $\widehat{\Psi }$ is the fermionic field operator, $\widehat{H}_0$ is
the one-particle Hamiltonian in Wiggler and interaction Hamiltonian is 
\begin{equation}
\widehat{H}_{int}=\frac ec\int \widehat{j}A_wd{\bf r}  \label{hint}
\end{equation}
with the current density operator 
\begin{equation}
\widehat{j}=\widehat{\Psi }^{+}\gamma _0\gamma \widehat{\Psi }  \label{cur}
\end{equation}
We pass to the Furry representation and write the Heisenberg field operator
of the electron in the form of an expansion over the stationary states of
type (\ref{11}) 
\begin{equation}
\widehat{\Psi }({\bf x},t)=\sum_{{\bf q},\sigma }\widehat{a}_{{\bf q},\sigma
}(t)\left| {\bf q},\sigma \right\rangle  \label{13}
\end{equation}
where we have excluded the antiparticle operators, since the contribution of
particle-antiparticle intermediate states will lead only to small
corrections to the processes considered. The creation and annihilation
operators $\widehat{a}_{{\bf q},\sigma }^{+}(t)$ and $\widehat{a}_{{\bf q}%
,\sigma }(t)$ associated with the positive energy solutions satisfy the
anticommutation rules at equal times 
\begin{eqnarray}
\{\widehat{a}_{{\bf q},\sigma }^{+}(t),\widehat{a}_{{\bf q}^{\prime },\sigma
^{\prime }}(t^{\prime })\}_{t=t^{\prime }} &=&\delta _{{\bf q},{\bf %
q^{\prime }}}\delta _{\sigma ,\sigma ^{\prime }}\;\;  \label{com} \\
\{\widehat{a}_{{\bf q},\sigma }^{+}(t),\widehat{a}_{{\bf q}^{\prime },\sigma
^{\prime }}^{+}(t^{\prime })\}_{t=t^{\prime }} &=&\{\widehat{a}_{{\bf q}%
,\sigma }(t),\widehat{a}_{{\bf q}^{\prime },\sigma ^{\prime }}(t^{\prime
})\}_{t=t^{\prime }}=0\;\;.  \nonumber
\end{eqnarray}

Taking into account (\ref{13}), (\ref{cur}), (\ref{hint}) and (\ref{11}) the
second quantized Hamiltonian can be expressed in the form 
\begin{eqnarray}
\widehat{H} &=&\sum\limits_{{\bf q},\sigma }\varepsilon ({\bf q})\widehat{a}%
_{{\bf q,}\sigma }^{+}\widehat{a}_{{\bf q,}\sigma }+\frac e{2c}\overline{A}%
_e\sum\limits_s\sum\limits_{{\bf q}_1,\sigma _3,\sigma _4}\widehat{a}_{{\bf q%
}_1-\hbar {\bf k+}s\hbar {\bf k}_0,\sigma _4}^{+}\widehat{a}_{{\bf q}%
_1,\sigma _3}\left\langle {\bf q}_1-\hbar {\bf k+}s\hbar {\bf k}_0,\sigma
_4\mid \mid s\mid \mid {\bf q}_1,\sigma _3\right\rangle e^{i\Delta ({\bf q}%
_1-\hbar {\bf k+}s\hbar {\bf k}_0,{\bf q}_1)t}  \nonumber \\
&&+\frac e{2c}\overline{A}_e^{*}\sum\limits_s\sum\limits_{{\bf q}_1,\sigma
_3,\sigma _4}\widehat{a}_{{\bf q}_1-\hbar {\bf k+}s\hbar {\bf k}_0,\sigma
_4}^{+}\widehat{a}_{{\bf q}_1,\sigma _3}\left\langle {\bf q}_1-\hbar {\bf k+}%
s\hbar {\bf k}_0,\sigma _4\mid \mid s\mid \mid {\bf q}_1,\sigma
_3\right\rangle e^{i\Delta ({\bf q}_1-\hbar {\bf k+}s\hbar {\bf k}_0,{\bf q}%
_1)t}  \nonumber \\
&&+\frac e{2c}\overline{A}_e^{*}\sum\limits_s\sum\limits_{{\bf q}_1,\sigma
_3,\sigma _4}\widehat{a}_{{\bf q}_1+\hbar {\bf k+}s\hbar {\bf k}_0,\sigma
_4}^{+}\widehat{a}_{{\bf q}_1,\sigma _3}\left\langle {\bf q}_1+\hbar {\bf k+}%
s\hbar {\bf k}_0,\sigma _4\mid \mid s\mid \mid {\bf q}_1,\sigma
_3\right\rangle e^{-i\Delta ({\bf q}_1,{\bf q}_1+\hbar {\bf k+}s\hbar {\bf k}%
_0,)t}.  \label{InH}
\end{eqnarray}
Here

\[
\left\langle {\bf q}_2,\sigma _2\mid \mid s\mid \mid {\bf q}_1,\sigma
_1\right\rangle =\frac{\overline{u}_{\sigma _2}(p_2)}{2\sqrt{\varepsilon
_1\varepsilon }}\left\{ \left( \widehat{e}_1-\frac{g^2e^2A_0^2(k_0e_1)%
\widehat{k}_0}{2c^2(p_1k_0)(p_2k_0)}\right) \Lambda _0\right. 
\]
\begin{equation}
\left. -eA_0\left( \frac{\gamma _x\widehat{k}_0\widehat{e}_1}{2c(p_1k_0)}+%
\frac{\widehat{e}_1\widehat{k}_0\gamma _x}{2c(p_2k_0)}\right) \Lambda
_1-eA_0\left( \frac{\gamma _y\widehat{k}_0\widehat{e}_1}{2c(p_1k_0)}+\frac{%
\widehat{e}_1\widehat{k}_0\gamma _y}{2c(p_2k_0)}\right) \Lambda _1^{\prime
}+(g^2-1)\frac{e^2A_0^2(k_0e_1)\widehat{k}_0}{2c^2(p_1k_0)(p_2k_0)}\Lambda
_2\right\} u_{\sigma _1}(p_1)  \label{1}
\end{equation}
where we have introduced the following functions \cite{Rit}

\begin{equation}
\{\sin \varphi ,\cos ^n\varphi \}\exp \left[ i\left( \alpha \sin (\varphi
-\varphi _0)-\beta \sin 2\varphi \right) \right] =\sum\limits_s\{\Lambda
_1^{\prime }(\alpha ,\beta ,s),\Lambda _n(\alpha ,\beta ,s)\}\exp (is\varphi
)  \label{2'}
\end{equation}
and parameters are defined as following 
\begin{equation}
\alpha =\frac{eA_0}{\hbar c}\left\{ \left( \frac{p_{1x}}{(p_1k_0)}-\frac{%
p_{2x}}{(p_2k_0)}\right) ^2+g^2\left( \frac{p_{1y}}{(p_1k_0)}-\frac{p_{2y}}{%
(p_2k_0)}\right) ^2\right\} ^{1/2},  \label{3'}
\end{equation}
\begin{equation}
\beta =(g^2-1)\frac{e^2A_0^2}{8c^2\hbar }\left( \frac 1{(p_1k_0)}-\frac 1{%
(p_2k_0)}\right) ,  \label{4'}
\end{equation}
\begin{equation}
\sin \varphi _0=\frac{eA_0}{\alpha \hbar c}g\left( \frac{p_{1y}}{(p_1k_0)}-%
\frac{p_{2y}}{(p_2k_0)}\right) ,  \label{5'}
\end{equation}
and

\begin{equation}
\Delta ({\bf q}_1-\hbar {\bf k+}s\hbar {\bf k}_0,{\bf q}_1)=\frac{%
\varepsilon ({\bf q}_1-\hbar {\bf k+}s\hbar {\bf k}_0)-\varepsilon ({\bf q}%
_1)+\hbar \omega }\hbar  \label{6'}
\end{equation}
is the resonance detuning.

We will use Heisenberg representation where operators evolution are given by
the following equation

\begin{equation}
i\hbar \frac{\partial \widehat{L}}{\partial t}=\left[ \widehat{L},\widehat{H}%
\right] ,  \label{ev}
\end{equation}
and expectation values are determined by the initial density matrix $%
\widehat{D}$

\begin{equation}
<\widehat{L}>=Sp\left( \widehat{D}\widehat{L}\right) .  \label{mean}
\end{equation}
The equations (\ref{ev}) should be supplemented by the Maxwell equation for $%
\overline{A}_e$ which is reduced to 
\begin{equation}
\frac{\partial A_e}{\partial t}+\frac{c^2{\bf k}}\omega \frac{\partial A_e}{%
\partial {\bf r}}=-i\frac{4\pi c}\omega \overline{<\widehat{j}e_1>\exp (ikx)}
\label{RM}
\end{equation}
where bar denotes averaging over time and space much larger than ($1/\omega
,1/k$) and 
\begin{equation}
<\widehat{j}e_1>=Sp\left( e_1\widehat{j}\widehat{D}\right)  \label{7'}
\end{equation}

\begin{equation}
e_1\widehat{j}=e\sum\limits_s\sum\limits_{{\bf q}_1,\sigma _1}\sum\limits_{%
{\bf q}_2,\sigma _2}\widehat{a}_{{\bf q}_2,\sigma _2}^{+}\widehat{a}_{{\bf q}%
_1,\sigma _1}\left\langle {\bf q}_2,\sigma _2\mid \mid s\mid \mid {\bf q}%
_1,\sigma _1\right\rangle e^{\frac i\hbar ({\bf q}_1-{\bf q}_2{\bf +}s\hbar 
{\bf k}_0){\bf r}}e^{\frac i\hbar (\varepsilon ({\bf q}_2)-\varepsilon ({\bf %
q}_1))t}  \label{8'}
\end{equation}
As we are interested in amplification of the wave with given $\omega ,{\bf k}
$, then we will keep only the resonant terms in (\ref{8'}) with ${\bf q}_2=%
{\bf q}_1-\hbar {\bf k+}s\hbar {\bf k}_0$. In principle, due to electron
beam energy and angular spreads different harmonics may contribute to the
process considered, but in the quantum regime (see below (\ref{25'}), (\ref
{26'}) ) we can keep only one harmonic ($s_0$). And for the resonant current
amplitude we will have the following expression

\begin{equation}
-i\overline{(e_1\widehat{j})\exp (ikx)}=\sum\limits_{{\bf q}}\widehat{\Pi }(%
{\bf q})  \label{9'}
\end{equation}
where 
\begin{equation}
\widehat{\Pi }({\bf q})=-ie\sum\limits_{\sigma _1,\sigma _2}\widehat{a}_{%
{\bf q}_{-},\sigma _2}^{+}\widehat{a}_{{\bf q},\sigma _1}\left\langle {\bf q}%
_{-},\sigma _2\mid \mid s_0\mid \mid {\bf q},\sigma _1\right\rangle
^{i\Delta ({\bf q}_{-},{\bf q})t}.  \label{10'}
\end{equation}
Here we have introduced the following notation

\begin{equation}
{\bf q}_{-}{\bf =q}-\hbar {\bf k+}s_0\hbar {\bf k}_0  \label{11'}
\end{equation}
The physical meaning of Eq. (\ref{10'}) is obvious: it describes the process
where a particle with quazimomentum ${\bf q}$ is annihilated and is created
in the state with the quazimomentum ${\bf q}-\hbar {\bf k+}s_0\hbar {\bf k}%
_0 $. Taking into account Eqs. \ref{InH}, \ref{ev}, \ref{com} for the
operator $\widehat{\Pi }({\bf q})$ we obtain the following equation

\begin{eqnarray}
\frac{\partial \widehat{\Pi }({\bf q})}{\partial t}-i\Delta ({\bf q}_{-},%
{\bf q}_1)\widehat{\Pi }({\bf q}) &=&\frac{e^2}{2\hbar c}\overline{A}%
_e\sum\limits_{\sigma _1,\sigma _2,\sigma _3}\left\{ \left\langle {\bf q}%
,\sigma _1\mid \mid -s_0\mid \mid {\bf q}_{-},\sigma _3\right\rangle
\left\langle {\bf q}_{-},\sigma _2\mid \mid s_0\mid \mid {\bf q},\sigma
_1\right\rangle \widehat{a}_{{\bf q}_{-},\sigma _2}^{+}\widehat{a}_{{\bf q}%
_{-},\sigma _3}\right.  \label{12'} \\
&&\left. -\left\langle {\bf q},\sigma _3\mid \mid -s_0\mid \mid {\bf q}%
_{-},\sigma _2\right\rangle \left\langle {\bf q}_{-},\sigma _2\mid \mid
s_0\mid \mid {\bf q},\sigma _1\right\rangle \widehat{a}_{{\bf q},\sigma
_3}^{+}\widehat{a}_{{\bf q},\sigma _1}\right\}  \nonumber
\end{eqnarray}
where we have kept only the resonant terms as far as these terms are
predominant in near-resonant emission/absorbtion, since their respective
detuning are much smaller than those for non-resonant terms, which are
detuned from the resonance due to $\omega >>\left| \Delta ({\bf q}_{-},{\bf q%
})\right| $.

We will assume that electron beam is non-polarized. This means that initial
one particle density matrix in the momentum space is 
\begin{equation}
\rho _{\sigma _1\sigma _2}({\bf q}_1,{\bf q}_2,0)=<\widehat{a}_{{\bf q}%
_2,\sigma _2}^{+}(0)\widehat{a}_{{\bf q}_1,\sigma _1}(0)>=\rho _0({\bf q}_1,%
{\bf q}_2)\delta _{\sigma _1,\sigma _2}.  \label{13'}
\end{equation}
Here $\rho _0({\bf q},{\bf q})$ is connected with the classical momentum
distribution function $n({\bf q})$ by the following formula 
\begin{equation}
\rho _0({\bf q},{\bf q})=\frac{(2\pi \hbar )^3}2n({\bf q}).  \label{14}
\end{equation}
For the expectation value of $\widehat{\Pi }({\bf q})$ from (\ref{12'}) we
will have

\begin{equation}
\frac{\partial \Pi ({\bf q})}{\partial t}-i\Delta ({\bf q}_{-},{\bf q}_1)\Pi
({\bf q})=\frac{e^2M^2}{2\hbar c}\overline{A}_e\left( \rho ({\bf q}_{-},{\bf %
q}_{-},t)-\rho ({\bf q},{\bf q},t)\right) ,  \label{15}
\end{equation}
where 
\[
\rho ({\bf q}_1,{\bf q}_1,t)=<\widehat{a}_{{\bf q}_1,\sigma _1}^{+}(t)%
\widehat{a}_{{\bf q}_1,\sigma _1}(t)> 
\]
\[
M^2=\sum\limits_{\sigma _1,\sigma _2}\left\langle {\bf q},\sigma _1\mid \mid
-s_0\mid \mid {\bf q}_{-},\sigma _2\right\rangle \left\langle {\bf q}%
_{-},\sigma _2\mid \mid s_0\mid \mid {\bf q},\sigma _1\right\rangle . 
\]
The quantity $M^2$ is reduced to the usual calculation of trace \cite{Lan}, 
\cite{Rit} and in our notations we have 
\begin{equation}
M^2=\frac{2c^4}{\varepsilon _{-}\varepsilon }\left| (pe_1^{\prime })\Lambda
_0+\frac{eA_0}c(e_{1x}^{\prime }\Lambda _1+ge_{1y}^{\prime }\Lambda
_1^{\prime })\right| ^2,  \label{16}
\end{equation}
where

\begin{equation}
e_1^{\prime }=e_1-k\left( \frac{ke_1}{k_0k}\right)  \label{17}
\end{equation}

Here we have neglected terms in order of $(\hbar \omega /\varepsilon )^2<<1$
as far as for the FEL this condition is always satisfied. Taking into
account Eqs.( \ref{InH}, \ref{ev}, \ref{com}) for the $\rho ({\bf q},{\bf q}%
,t)$ and $\rho ({\bf q}_{-},{\bf q}_{-},t)$ we will obtain

\begin{equation}
\frac{\partial \rho ({\bf q},{\bf q},t)}{\partial t}=\frac 1{4\hbar c}\left(
A_e^{*}\Pi +A_e\Pi ^{*}\right)  \label{18}
\end{equation}
\begin{equation}
\frac{\partial \rho ({\bf q}_{-},{\bf q}_{-},t)}{\partial t}=-\frac 1{4\hbar
c}\left( A_e^{*}\Pi +A_e\Pi ^{*}\right)  \label{19}
\end{equation}

To take into account the pulse propagation effects we can replace the time
derivatives by the following expression

\[
\frac \partial {\partial t}\rightarrow \frac \partial {\partial t}+\overline{%
{\bf v}}\frac \partial {\partial {\bf r}}, 
\]
where $\overline{{\bf v}}$ is the mean velocity of electron beam and
convectional part of derivative expresses the pulse propagation effects.
Introducing new variables 
\begin{equation}
\delta n=\frac 2{(2\pi \hbar )^3}\left[ \rho ({\bf q}_{-},{\bf q}%
_{-},t)-\rho ({\bf q},{\bf q},t)\right]  \label{20}
\end{equation}
\begin{equation}
\frac 1{(2\pi \hbar )^3}\Pi ({\bf q})=J({\bf q})  \label{21}
\end{equation}
and replacing summation in (\ref{RM}) by integration, the self-consistent
set of equations reads 
\[
\frac{\partial J({\bf q})}{\partial t}+\overline{{\bf v}}\frac{\partial J(%
{\bf q})}{\partial {\bf r}}-i\Delta J({\bf q})=\frac{e^2M^2}{4\hbar c}%
A_e\left( x,z,t\right) \delta n({\bf q}) 
\]
\begin{equation}
\frac{\partial \delta n({\bf q})}{\partial t}+\overline{{\bf v}}\frac{%
\partial \delta n({\bf q})}{\partial {\bf r}}=-\frac 1{\hbar c}\left(
A_e^{*}J({\bf q})+A_eJ^{*}({\bf q})\right)  \label{main}
\end{equation}
\[
\frac{\partial A_e}{\partial t}+\frac{c^2{\bf k}}\omega \frac{\partial A_e}{%
\partial {\bf r}}=\frac{4\pi c}\omega \int d{\bf q}J({\bf q}). 
\]
These equations yield to the conservation laws for the energy of the system
and particle number: 
\begin{equation}
\frac{\partial \mid A_e\mid ^2}{\partial t}+\frac{c^2{\bf k}}\omega \frac{%
\partial \mid A_e\mid ^2}{\partial {\bf r}}=-\frac{4\pi \hbar c^2}\omega
\int d{\bf q}\left( \frac \partial {\partial t}+\overline{{\bf v}}\frac %
\partial {\partial {\bf r}}\right) \delta n({\bf q})  \label{22}
\end{equation}
\begin{equation}
\left( \frac \partial {\partial t}+\overline{{\bf v}}\frac{\partial )}{%
\partial {\bf r}}\right) \left( \delta n({\bf q})^2+\frac 8{e^2M^2}\left| J(%
{\bf q})\right| ^2\right) =0  \label{22'}
\end{equation}

From the set of equations (\ref{main}) by the perturbation theory one can
obtain the small signal gain, which in the quasiclassical limit will
coincide with the classical one.

The emission and absorbtion are characterized by the following widths

\begin{eqnarray}
\Delta _e &=&s\omega _0\frac vc\cos \theta _1-\omega (1-\frac vc\cos \theta
)-\frac{ce^2A_0^2\omega ^{\prime }}{2\varepsilon ^2v\cos \theta _1}\cos
\theta _0  \label{23} \\
&&-\frac{s_0\hbar \omega _0\omega }\varepsilon \cos \theta _0  \nonumber
\end{eqnarray}

\begin{equation}
\Delta _a=\Delta _e+\frac{2s_0\hbar \omega _0\omega }\varepsilon \cos \theta
_0,  \label{24}
\end{equation}
where $\omega _0=2\pi c/\ell $ , $\theta $ and $\theta _0$ are the
scattering angles of probe photons with respect to the electron beam
direction of motion and undulator axis, respectively, $\theta _1$ is the
angle of the electron beam direction of motion with respect to undulator
axis.

The quantum regime assumes

\begin{equation}
\Delta _a-\Delta _e=\frac{2s\hbar \omega _0\omega }\varepsilon \cos \theta
_0>\max \left\{ \left| \frac{\partial \Delta _e}{\partial \eta _i}\delta
\eta _i+\frac{\partial ^2\Delta _e}{\partial \eta _i^2}(\delta \eta
_i)^2\right| ,\frac{\omega _0}N\right\} ,  \label{*con}
\end{equation}
where by $\eta _i$ we denote the set of quantities characterizing electron
beam and pump field and by $\delta \eta _i$ their spreads. The second term
in the figured brackets of Eq. (\ref{*con}) expresses resonance width caused
by finite interaction length, $N$ is the number of periods of pump field.

Particularly for energetic ($\Delta \varepsilon $) and angular ($\Delta
\vartheta $) spreads from Eq. (\ref{*con}) (for $\theta _0,\theta <<1$) we
will have

\begin{equation}
\Delta \varepsilon \prec \hbar \omega  \label{25}
\end{equation}
\begin{equation}
\left| \theta \Delta \vartheta +\frac{\Delta \vartheta ^2}2\right| <\frac{%
2s_0\hbar \omega _0}\varepsilon  \label{26}
\end{equation}

The conditions of keeping only one harmonic in resonant current, are:

\begin{equation}
\frac{\Delta \varepsilon }\varepsilon <<1/s_0  \label{25'}
\end{equation}
\begin{equation}
\left| \theta \Delta \vartheta +\frac{\Delta \vartheta ^2}2\right| <\frac{%
\omega _0}\omega  \label{26'}
\end{equation}
As we see, these conditions are weaker than the conditions of quantum
regimes (\ref{25}) and are well enough satisfied for actual beams.

\section{Steady-State Regimes of Amplification}

Our goal is to determine the conditions under which we will have non-linear
amplification. We will assume steady-state operation, according to which
will drop with partial time derivatives in (\ref{main}). The considered
setup is either a single-pass amplifier, for which it is necessary injected
input signal or self-amplified coherent spontaneous emission, for which it
is necessary initially modulated beam.

Besides we will consider the case of exact resonance neglecting detuning in
Eqs. (\ref{main}) assuming that electron beam momentum distribution is
centered at $\Delta _e=0$. To achieve maximal Doppler-shift and optimal
conditions of amplification we will assume that electron beam propagates
along undulator axis ($Z$ axis). In this case optimal condition for LW is $%
\theta =0$, while for HW $\theta \sim K/\gamma _L$ ($\theta <<1$). For both
cases we will assume that the envelope of probe wave depends only on $z$ .
Then the set of equations (\ref{main}) and conservation laws (\ref{22}, \ref
{22'}) are reduced to 
\[
\frac{\partial J}{\partial z}=\frac{e^2M^2}{4\hbar c\overline{v}_z}A_0\Delta
n 
\]
\[
\frac{\partial \delta n}{\partial z}=-\frac 2{\hbar c\overline{v}_z}A_0J 
\]
\begin{equation}
\frac{\partial A_0}{\partial z}=\frac{4\pi }\omega J  \label{27}
\end{equation}
\[
\delta n^2+\frac 8{e^2M^2}\mid \Pi \mid ^2=N_0^2 
\]
\[
I=I_0+\frac{\hbar \omega \overline{v}_z}2\left( \delta n_0-\delta n\right) , 
\]
where $N_0$ is the beam density , $I$ is the probe wave intensity and $I_o$
is the initial one. From Eq.(\ref{27}) we have the following expressions for 
$J$ and $\delta n$ : 
\[
\delta n=N_0\cos \left\{ \frac{e\left| M\right| }{2^{1/2}\hbar c\overline{v}%
_z}\int_0^zA_0dz+\varphi _o\right\} 
\]
\begin{equation}
J=\frac{e\left| M\right| }{2^{3/2}}N_0\sin \left\{ \frac{e\left| M\right| }{%
2^{1/2}\hbar c\overline{v}_z}\int_0^zA_0dz+\varphi _o\right\} ,  \label{28}
\end{equation}
where $\varphi _o$ is determined by boundary conditions. Denoting 
\begin{equation}
\varphi =\frac{e\left| M\right| }{2^{1/2}\hbar c\overline{v}_z}%
\int_0^zA_0dz+\varphi _o  \label{29}
\end{equation}
we arrive to the nonlinear pendulum equation 
\begin{equation}
\frac{\partial ^2\varphi }{\partial z^2}=\sigma ^2\sin \varphi ,  \label{30}
\end{equation}
where 
\begin{equation}
\sigma ^2=\frac{\pi e^2M^2N_0}{\hbar \omega c\overline{v}_z}  \label{31}
\end{equation}
is the main characteristic parameter of amplification: $L_c=1/\sigma $ is
the characteristic length of amplification. For the LW from Eqs.(\ref{2'}), (%
\ref{3'}), (\ref{4'}), (\ref{5'}) and (\ref{16}) we have

\begin{equation}
\sigma _L=\frac{K\Lambda _1(0,\beta ,s_0)}{\gamma _L^2}\sqrt{\alpha _0\frac{%
c\ell }{2s_0\overline{v}_z}N_0(1+K^2/2)}  \label{32}
\end{equation}
where $\alpha _0$ is the fine structure constant and the function $\Lambda
_1(0,\beta ,s)$ is expressed by the ordinary Bessel functions: 
\begin{equation}
\Lambda _1(0,\beta ,s_0)\simeq \frac 12\left| J_{\frac{s_0-1}2}\left( \frac{%
s_0K^2}{4+2K^2}\right) -J_{\frac{s_0+1}2}\left( \frac{s_0K^2}{4+2K^2}\right)
\right| .  \label{33}
\end{equation}
In this case only odd harmonics are possible. For the HW we have

\begin{equation}
\sigma _L=\frac K{\gamma _L^2}\left( \frac{\theta \gamma _L}K+\frac{s_0}%
\alpha \right) \left| J_{s_0}\left( \alpha \right) \right| \sqrt{\alpha _0%
\frac{c\ell }{2s_0\overline{v}_z}N_0(1+K^2+\theta ^2\gamma _L^2)}  \label{34}
\end{equation}
and the argument of Bessel function is

\begin{equation}
\alpha \simeq \frac{2s_0K\gamma _L\theta }{1+K^2+\theta ^2\gamma _L^2}.
\label{35}
\end{equation}

We will consider two regimes of amplification, which are determined by
initial conditions. For the first regime the initial macroscopic transition
current of the electron beam is zero and it is necessary to have an seeding
electromagnetic wave. In this case the following boundary conditions are
imposed 
\begin{equation}
\delta n\mid _{z=0}=N_0;\quad J\mid _{z=0}=0;\quad I\mid _{z=0}=I_0
\label{b1}
\end{equation}
The solution of Eq.(\ref{21}) in this case reads 
\begin{equation}
I\left( z\right) =\frac{I_0}{dn^2\left( \frac \sigma \kappa z;\kappa \right) 
}  \label{b2}
\end{equation}
\begin{equation}
\kappa =\left( 1+\frac{I_0}{N_0\hbar \omega v_z}\right) ^{-\frac 12}
\label{m1}
\end{equation}
where $dn\left( z,\kappa \right) $ is the Elliptic function of Jacobi and $%
\kappa $ its module.

As is known $dn\left( z,\kappa \right) $ is the periodic function with the
period $2\Xi (\kappa )$ , where $\Xi (\kappa )$ is the Complete Elliptic
Integral of first order. At the distances $L=(2r+1)\kappa \cdot \Xi (\kappa
)/\sigma $ ($r=0,1,2...$) the wave intensity reaches its maximal value which
equals to 
\begin{equation}
I_{\max }=I_0+N_0\hbar \omega v_z.  \label{max1}
\end{equation}
For the short interaction lengths $z\ll L_c$ from Eq.(\ref{21}) we have 
\[
I\left( z\right) =I_0\left( 1+\sigma ^2z^2\right) 
\]
and the wave gain is rather small. To extract maximal energy from electron
beam the interaction length should be at least order of half of spatial
period of the wave envelope variation- $\kappa \cdot \Xi (\kappa )/\sigma $.
At this condition the intensity value $I_{\max }=I_0+N\hbar \omega v_z$ is
achieved, because all electrons make contribution in radiation field. Taking
into account that seed power is much more smaller than $I_{\max }$ and when $%
1-\kappa <<1$

\[
\Xi (\kappa )\rightarrow \frac 12\ln \left[ \frac{16}{1-\kappa ^2}\right] 
\]
for amplification length we will have 
\begin{equation}
L\simeq L_c\left( 1.4+\ln \left( I_{\max }/I_0\right) \right) .  \label{36}
\end{equation}
Let us now consider the other regime of wave amplification when electron
beam is modulated- ''macroscopic transition current'' $J$ differs from zero.
This regime can operate without any initial seeding power ($I_0=0$). So we
will consider the optimal case with the following initial conditions 
\begin{equation}
J\mid _{z=0}=J_0\qquad \delta n\mid _{z=0}=\delta n_0;\qquad I\mid _{z=0}=0
\label{b3}
\end{equation}
Then the wave intensity is expressed by the formula 
\begin{equation}
I\left( z\right) =\frac{N_0\hbar \omega v_z}2\left( 1-\frac{\Delta n_0}{N_0}%
\right) \left[ \frac 1{dn^2(\ae z;k)}-1\right]  \label{b4}
\end{equation}
and module is

\begin{equation}
\kappa ^2=\frac 12(1+\frac{\Delta n_0}{N_0})  \label{37}
\end{equation}
As is seen from (\ref{b4}) in this case the intensity varies periodically
with the distances as well, with the maximal value of intensity 
\begin{equation}
I_{\max }=\frac{N_0\hbar \omega v_z}2(1+\frac{\Delta n_0}{N_0}).  \label{38}
\end{equation}

The second regime is more interesting. It is the regime of amplification
without initial seeding power and has superradiant nature. For the short
interaction length $z\ll L_c$ according to (\ref{b4}) 
\begin{equation}
I\left( z\right) =\frac{N_0\hbar \omega v_z\sigma ^2z^2}4\left( 1-\frac{%
\Delta n_0}{N_0}\right)  \label{39}
\end{equation}
The intensity is scaled as $N_0^2$, which means that we have superradiation
The radiation intensity in this regime reaches significant value even at $%
z\ll L$ .

\section{Discussion}

The coherent interaction time of electrons with probe radiation is confined
by the several relaxation processes. To be more precise in the self
consistent set of equations (\ref{main}) we should add the terms describing
spontaneous transitions and other relaxation processes. Since we have not
taken into account the relaxation processes, this consideration is correct
only for the distances $L\preceq c\tau _{\min }$, where $\tau _{\min }$ is
the minimum of all relaxation times. Due to spontaneous radiation electron
will lose the energy $\sim \hbar \omega $ at the distances

\begin{equation}
L_s\simeq c\frac{\hbar \omega }{I_s}=\frac 3{2\pi }\frac{s_0\ell }{\alpha
_0(1+K^2/2)K^2},  \label{s1}
\end{equation}
where $I_s$ is the intensity of spontaneous radiation (this is for LW; for
HW one should replace $K^2\rightarrow 2$ $K^2$). Although the cutoff
harmonic is increased $s_c\sim $ $K^3$ with increasing of $K$ for the large
wiggler parameter $K$ $\succsim 1$ the rule of spontaneous radiation is
increased $L_s\sim K^{-4}$ and above mentioned regimes will be interrupted.
The obtained solutions are correct for the distances $\sim L_s$.

In the Tables 1, 2 we give the parameters for the different setups of a beam
and undulator as for LW as well as for HW. The beam current have been chosen 
$5kA$ and the beam radius $10^{-3}cm$. Maximum intensity $I_{\max }$ scales
as $\sim (\hbar \omega /mc^2)\cdot 7\cdot 10^{14}\cdot W/cm^2$. As we see
from these tables for the high harmonics $L_c$ is decreased and
simultaneously $\hbar \omega /\varepsilon $ is increased, but $L_s<L_c.$ In
this case wiggler tapering is necessary to keep the resonance condition. But
how it will look like for quantum regime will be the subject of the future
work.

The second regime may be more promising as it allows considerable output
intensities even for small interaction lengths. For the $s=51$ (Table 1. $%
\hbar \omega \sim 1.1Mev$) at the $L=L_s$ ($\sigma ^2z^2<<1$), $1-\delta
n_0/N_0\simeq 10^{-2}$ ($1\%$ modulation) from (\ref{39}) we have $I$ $\sim
10^{10}W/cm^2.$ It is expected that the effects of energy and angular
spreads will not have significant influence upon this regime as it is
governed by the initial current and only Doppler dephasing and spontaneous
lifetime may interrupt the superradiation process. Note, that necessary for
this regime quantum modulation of the particle beams at the above optical
frequencies can be obtained through multiphoton transitions in the laser
field at the presence of a ''third body''. The possibilities of quantum
modulation at hard X-ray frequencies in the induced Compton, undulator and
Cherenkov processes have been studied in the works \cite{Avet}.

This work is supported by International Science and Technology Center (ISTC)
Project No. A-353.

{\bf Table1 (LW)}

\begin{center}
\begin{tabular}{|l|l|l|l|l|l|l|}
\hline
$s$ & $\gamma _L$ & $K$ & $\ell [cm]$ & $\hbar \omega /\varepsilon $ & $%
L_c[m]$ & $L_s/L_c$ \\ \hline
$1$ & $5\cdot 10^4$ & $0.1$ & $1.5$ & $1.6\cdot 10^{-5}$ & $12$ & $8.5$ \\ 
\hline
$1$ & $10^5$ & $0.1$ & $1.5$ & $3.2\cdot 10^{-5}$ & $47.7$ & $2.14$ \\ \hline
$3$ & $4\cdot 10^4$ & $1$ & $3$ & $1.3\cdot 10^{-5}$ & $2.8$ & $1.1$ \\ 
\hline
$9$ & $2\cdot 10^4$ & $1.5$ & $4$ & $10^{-5}$ & $2$ & $1.5$ \\ \hline
$21$ & $2.5\cdot 10^4$ & $2.7$ & $5$ & $10^{-5}$ & $1.2$ & $0.7$ \\ \hline
$51$ & $5\cdot 10^4$ & $3$ & $5$ & $4.5\cdot 10^{-5}$ & $24$ & $0.05$ \\ 
\hline
\end{tabular}
\end{center}

{\bf Table 2. (HW)}

\medskip

\begin{center}
\begin{tabular}{|l|l|l|l|l|l|l|}
\hline
$s$ & $\gamma _L$ & $K$ & $\ell [cm]$ & $\hbar \omega /\varepsilon $ & $%
L_c[m]$ & $L_s/L_c$ \\ \hline
$1$ & $5\cdot 10^4$ & $0.07$ & $1.5$ & $1.6\cdot 10^{-5}$ & $16.8$ & $12.3$
\\ \hline
$1$ & $10^5$ & $0.2$ & $1.5$ & $3\cdot 10^{-5}$ & $21.5$ & $1.1$ \\ \hline
$3$ & $6\cdot 10^4$ & $1$ & $3$ & $10^{-5}$ & $1.9$ & $1$ \\ \hline
$10$ & $2\cdot 10^4$ & $1.05$ & $3$ & $10^{-5}$ & $2.2$ & $2.58$ \\ \hline
$20$ & $4\cdot 10^4$ & $1.8$ & $5$ & $10^{-5}$ & $2$ & $1.4$ \\ \hline
$51$ & $5\cdot 10^4$ & $2.3$ & $5$ & $2\cdot 10^{-5}$ & $5.8$ & $0.5$ \\ 
\hline
\end{tabular}
\end{center}

\end{document}